\documentclass[twocolumn,showpacs,preprintnumbers,prl,aps,amssymb,superscriptaddress]{revtex4} 
\usepackage{graphicx} 
\usepackage{dcolumn} 
\usepackage{bm} 
\begin{document} 
 
\newcommand{\Ir}{CeIrIn$_5$} 
\newcommand{\ie}{{\it i.e.}} 
\newcommand{\eg}{{\it e.g.}} 
\newcommand{\etal}{{\it et al.}} 
%%%%%%%%%%%%%%%%%%%%%%%%%%%% TITLE 
\title{Hybrid gap structure in the heavy-fermion superconductor CeIrIn$_5$} 
 
%%%%%%%%%%%%%%%%%%%%%%%%%%%% AUTHORS 
\author{H.~Shakeripour} 
\affiliation{D\'epartement de physique and RQMP, Universit\'e de Sherbrooke, Sherbrooke, Canada}  
 
\author{M.~A.~Tanatar} 
\altaffiliation[Permanent address: ]{Inst. Surface Chemistry, N.A.S. Ukraine, Kyiv, Ukraine.}
\affiliation{D\'epartement de physique and RQMP, Universit\'e de Sherbrooke, Sherbrooke, Canada}  
 
\author{S.~Y.~Li} 
\affiliation{D\'epartement de physique and RQMP, Universit\'e de Sherbrooke, Sherbrooke, Canada}  
 
\author{Louis Taillefer}  
\email{Louis.Taillefer@USherbrooke.ca}  
\affiliation{D\'epartement de physique and RQMP, Universit\'e de Sherbrooke, Sherbrooke, Canada}  
\affiliation{Canadian Institute for Advanced Research, Toronto, Ontario, Canada}  
 
\author{C.~Petrovic}  
\affiliation{Department of Physics, Brookhaven National Laboratory, Upton, New York 11973, USA}  
 
\date{\today} 
%%%%%%%%%%%%%%%%%%%%%%%%%%%% ABSTRACT 
\begin{abstract} 
 
The thermal conductivity $\kappa$ of the heavy-fermion superconductor CeIrIn$_5$ was measured as a function of temperature down to $T_c$/8, for current directions perpendicular ($J \parallel a$) and parallel ($J \parallel c$) to the tetragonal $c$ axis. For $J \parallel a$, a sizable residual linear term $\kappa_0 / T$ is observed, as previously, which confirms the presence of line nodes in the superconducting gap. For $J \parallel c$, on the other hand, $\kappa / T \to 0$ as $T \to 0$. The resulting precipitous decline in the anisotropy ratio $\kappa_c / \kappa_a$ at low temperature rules out a gap structure with line nodes running along the $c$-axis, such as the $d$-wave state favoured for CeCoIn$_5$, and instead points to a hybrid gap of $E_g$ symmetry.  
It therefore appears that two distinct superconducting states are realized in the Ce$M$In$_5$ family.  

\end{abstract} 
\pacs{74.70.Tx, 74.20.Rp, 74.25.Fy} 
\maketitle 
 
%%%%%%%%%%%%%%%%%%%%%%%%%%%% INTRODUCTION 
The discovery of magnetically-mediated superconductivity in the heavy-fermion material CeIn$_3$ \cite{Mathur} has attracted considerable attention as a possible archetype for unconventional pairing in a variety of superconductors. 
However, the fact that the superconducting state in this material only exists under pressure makes it difficult to know its actual pairing state.
Fortunately, the closely related family of Ce$M$In$_5$ compounds offers an ideal testing ground for investigating the role that dimensionality, magnetic order and fluctuations play in determining the strength and symmetry of the superconducting state, as two members of the family show superconducting order at ambient pressure ($M$ = Co, Ir) and the third shows antiferromagnetic order ($M$ = Rh) \cite{Petrovic,Petrovic-Ir}.  
In CeCoIn$_5$, the observation of a four-fold anisotropy in the thermal conductivity \cite{IzawaCo} and specific heat \cite{Aoki} on rotation of a magnetic field in the basal tetragonal plane points to a $d$-wave gap (presumably of $d_{x^2-y^2}$ symmetry \cite{Vekhter}). A number of theoretical models propose a $d_{x^{2}-y^{2}}$ state \cite{Takimoto,Watanabe,Tanaka}, analogous to that realized in cuprate superconductors. 
Since the calculated band structure \cite{Maehira} and measured Fermi surface \cite{Haga} of CeCoIn$_5$ and CeIrIn$_5$ are very similar, and properties like the specific heat \cite{Movshovich} and the NQR relaxation rate \cite{NQR,NQR2} exhibit the same temperature dependence, it has generally been assumed that the two superconductors have the same pairing state, even though their transition temperature $T_c$ differs by a factor of 6. However, because recent evidence suggests that the phase diagram of Ce$M$In$_5$ may contain more than one superconducting state \cite{Nicklas,Kawasaki}, it has now become crucial to pin down the pairing state of CeIrIn$_5$.  
 
One of the most conclusive ways to determine the pairing symmetry of a superconductor is to map out its gap structure. A powerful approach to probe the gap structure and locate the position of nodes around the Fermi surface is to measure quasiparticle heat transport as a function of direction, at very low temperature.  
As an example, the anisotropy of heat transport played a decisive role in elucidating the pairing symmetry of the hexagonal heavy-fermion superconductor UPt$_3$ ($T_c = 0.5$~K) \cite{Lussier,RMP-UPt3}. In this Letter, we report a study of heat transport in CeIrIn$_5$ ($T_c = 0.4$~K) down to $T_c / 8$ for current directions parallel and perpendicular to the tetragonal axis of a single crystal. It reveals a dramatic anisotropy as $T \to 0$, whereby low-energy nodal quasiparticles carry heat well in the basal plane but poorly, if at all, along the $c$-axis. This is inconsistent with the $d$-wave states proposed for CeCoIn$_5$, characterized by line nodes running along the $c$-axis. In fact, it eliminates all allowed (spin singlet) pairing symmetries but one, the $(1,i)$ state of the $E_{g}$ representation. This state has a hybrid gap structure, with a line node in the basal plane and point nodes in the $c$-direction, of the kind also found in UPt$_3$ \cite{RMP-UPt3}. 
%%%%%%%%%%%%%%%%%%%%%%%%%%%% EXPERIMENTAL 
 
Single crystals of CeIrIn$_5$ were grown by the self-flux method \cite{Petrovic-Ir}. Two samples were cut into parallelepipeds with dimensions $\sim 4.5 \times 0.14 \times 0.045$~mm$^3$ (for $J \parallel a$) and $\sim 1 \times 0.15 \times 0.086$~mm$^3$ (for $J \parallel c$). 
Their exceptionally low residual resistivity (at $T\to 0$ and $H\to 0$) attests to their very high purity: $\rho_{0a}~(\rho_{0c}) = 0.2~(0.5)~\mu\Omega$~cm. The bulk transition temperature is $T_c = 0.38 \pm 0.02$~K and the upper critical field $H_{c2} = 0.49$~T for $H \parallel c$. 
The thermal conductivity was measured in a dilution refrigerator using a standard four-wire steady-state method with two RuO$_2$ chip thermometers calibrated {\it in situ} against a reference Ge thermometer. The same indium contacts were used for electrical resistivity and thermal conductivity. Their typical resistance at low temperature was $\sim5$~m$\Omega$. Note that the contribution of phonons to the thermal transport is entirely negligible below 1~K. 
 
\begin{figure} 
\centerline{ 
\scalebox{0.49}{ 
\includegraphics{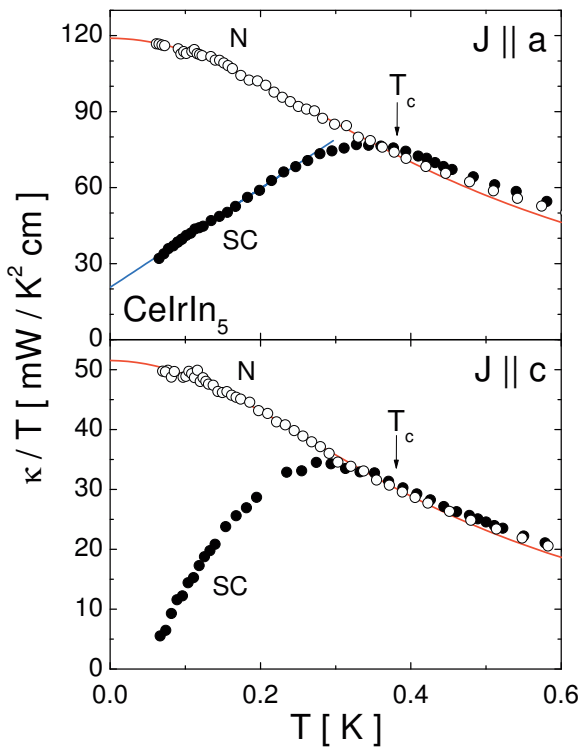}}} 
\caption{\label{fig.kT} 
Thermal conductivity $\kappa$ of CeIrIn$_5$, plotted as $\kappa/T$ vs $T$, for a heat current perpendicular (top) and parallel (bottom) to the $c$-axis, in the superconducting state (SC; $H=0$) and the normal state (N; $H=0.5$~T). The normal state data are fitted to the Fermi-liquid dependence $\kappa_N / T = 1 / (a+bT^2)$ (red line). A linear fit to the superconducting state data for $J \parallel a$ (below $T_c / 2$) is also shown (blue line). }  
\vglue-3mm  
\end{figure}

%%%%%%%%%%%%%%%%%%%%%%%%%  NORMAL STATE 
\textit{Normal state}. 
The thermal conductivity of CeIrIn$_5$ is plotted in Fig.~\ref{fig.kT} as $\kappa/T$ vs $T$, for a current perpendicular ($J \parallel a$) and parallel ($J \parallel c$) to the $c$ axis. We first concentrate on the normal state, where the electrical resistivity $\rho(T)$ was found to satisfy the Wiedemann-Franz law to better than 1~\%, as $T \to 0$: $\kappa_{N} / T = L_{0} / \rho_{0}$, where $L_{0}=\frac{\pi^{2}}{3}(\frac{k_{B}}{e})^{2}$. This shows that our measurements do not suffer from electron-phonon decoupling (see discussion in \cite{Msmith,Makariy}). $\kappa_N$ exhibits the temperature dependence of a Fermi liquid, $\kappa_{N}(T)/T = 1 / (a + bT^{2})$, with $a$ = 8.5 (19.6)~K$^{2}$~cm/W and $b$ = 36 (90)~cm/W, for $J\parallel a$ ($J\parallel c$).  
The fact that both samples have the same (thermal) resistivity ratio, namely $\kappa/T(T \to 0) / \kappa/T(0.6~{\rm K})$~= 2.4~(2.6) for $J\parallel a$ ($J\parallel c$), shows that they have the same level of impurity scattering.  
%--------------- figure 1: k/T Vs. T 
%------------------------------------------------------------------------ 
  
%%%%%%%%%%%%%%%%%%%%%%%%%  SUPERCONDUCTING STATE 
\textit{Superconducting state}. 
Given this conventional normal state behaviour, let us turn to the superconducting state, starting with $J \parallel a$. In the top panel of Fig.~\ref{fig.kT}, $\kappa_a / T$ is seen to be roughly linear in $T$, extrapolating to a finite residual linear term as $T \to 0$, as found by Movshovich {\it et al.} \cite{Movshovich}. This is the behaviour expected of a superconductor with line nodes, {\it i.e.} with a density of states that grows linearly with energy ($N(E) \propto E$) \cite{Fledderjohann,Graf,Norman}. 
In particular, theory shows that, in the $T \to 0$ limit, $\kappa/T$ reaches a universal value given by \cite{Graf,Graf-JLTP}: 
%-------------- 
\begin{eqnarray} 
\frac{\kappa_{0}}{T}=\frac{1}{3}~ \gamma_{N}~v^2_{F}~\frac{a\hbar}{2\mu\Delta_{0}}~~~, 
\end{eqnarray} 
%-------------- 
where $\gamma_N$ is the linear term in the normal state specific heat, $v_F$ is the Fermi velocity, $\Delta_0$ is the gap maximum, $\mu$ is the slope of the gap at the node, and $a$ is a constant of order unity whose value depends on the particular gap structure \cite{Graf}. $\kappa_{0}/T$ is called ``universal'' because it does not depend on impurity concentration and can therefore be used to measure the magnitude of the gap. Experimentally, universal conduction has been observed in high-$T_c$ cuprates \cite{Taillefer} and in the spin-triplet superconductor Sr$_2$RuO$_4$ \cite{Suzuki}, and Eq.~1 works well in both cases.  
 
\begin{figure} 
\centerline{ 
%\scalebox{0.22}{ 
\includegraphics[scale=0.168]{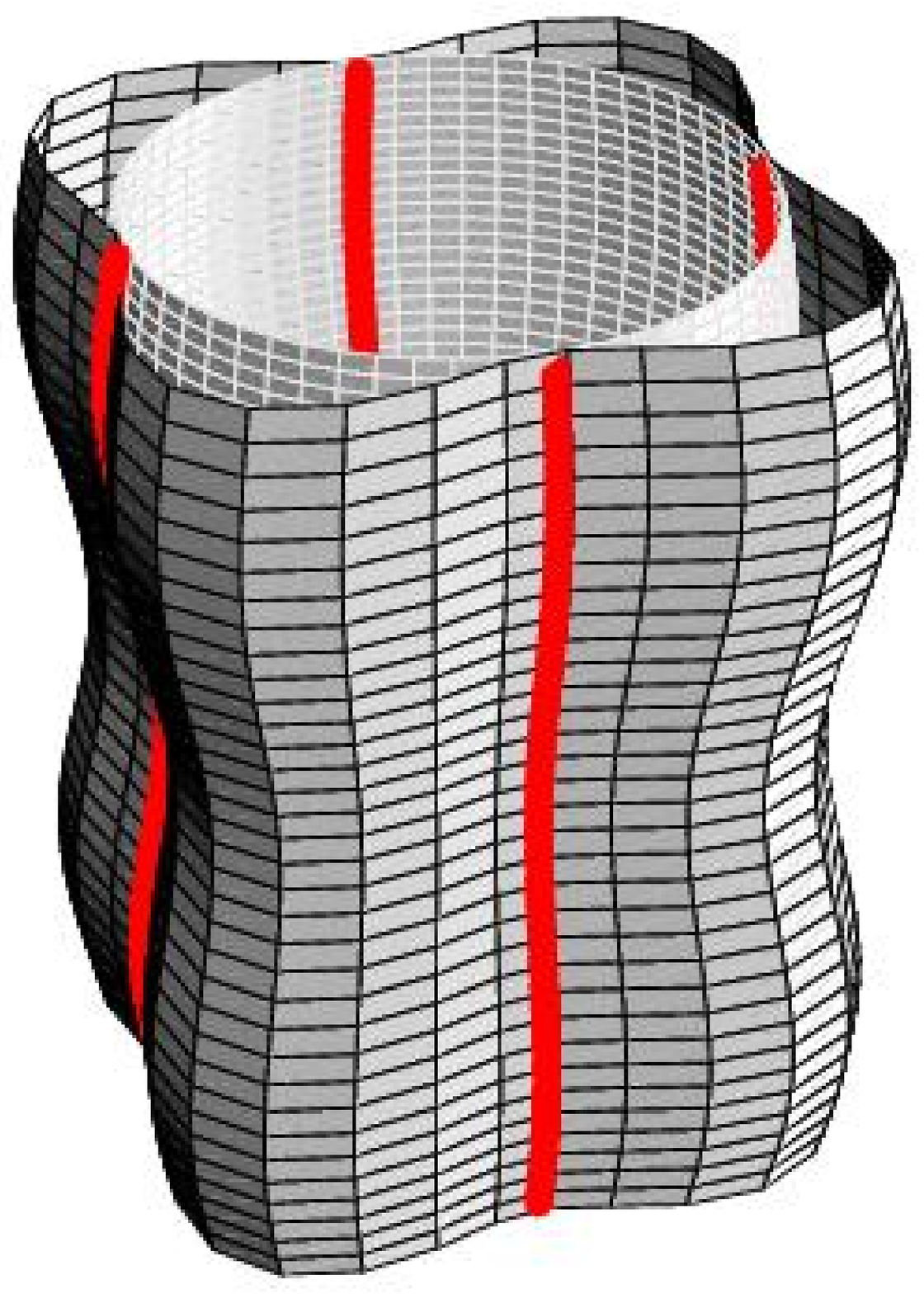} 
\includegraphics[scale=0.18]{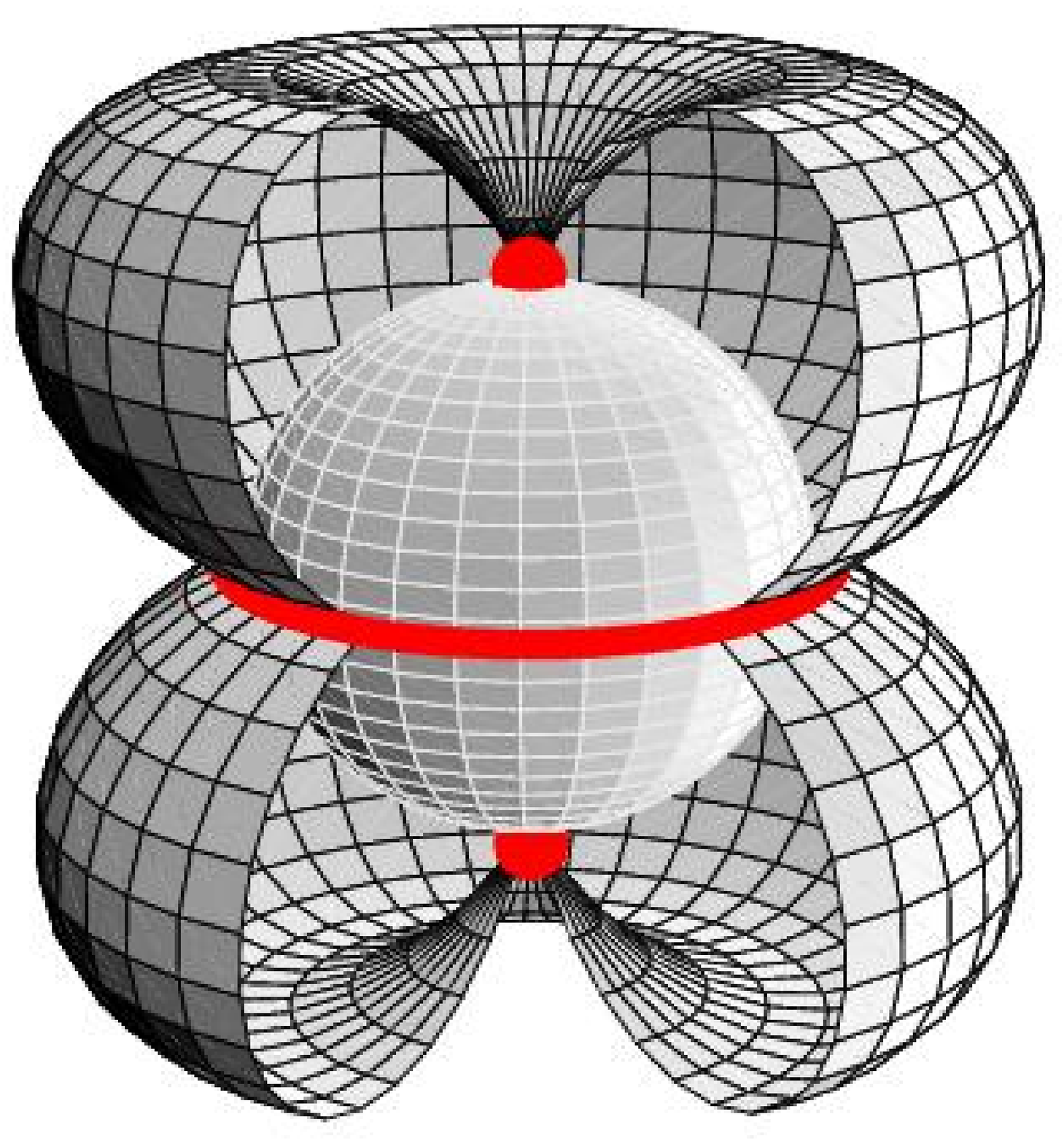}} 
\begin{picture}(0,0) 
\put(-50,1){\makebox(0,0){d-wave}} 
\put(48,1){\makebox(0,0){hybrid}} 
\put(-103,20){\makebox(0,0){c-axis}} 
\put(-105,25){\linethickness{0.25mm}\vector(0,1){25}} 
\end{picture} 
\caption{\label{fig:gaps} 
Typical gap structures in tetragonal symmetry, drawn on simple Fermi surfaces.  
{\it Left} : $d_{x^2 - y^2}$ gap on a (warped) cylindrical Fermi surface, with four vertical line nodes running along the $c$-axis. 
{\it Right} : hybrid gap on a spherical Fermi surface, with one horizontal line node in the basal plane and two point nodes along the $c$-axis. 
Quasiparticle heat conduction at low temperature is entirely governed by nodal topology.} 
\vglue-3mm 
\end{figure}

%--------------------------------------- Table 
\begin{table} 
\caption{Even-parity (spin-singlet) pair states in a tetragonal crystal with point group $D_{4h}$ \cite{D4h}. (V = vertical line node, H = horizontal line node.)} 
\vspace{.1cm}  
\begin{tabular*}{8.5cm}{lcccccc}\hline 
\vspace{.1cm}  
Representation &  & Gap & ~~~~&  Basis function  &~~~~~ & Nodes \\  \hline\ 
\\  
\vspace{.1cm}  
$A_{1g}$  & &s-wave & & 1, ($x^2+y^2$), $z^2$ & & none \\  
\vspace{.1cm}  
$A_{2g}$  & &g-wave  & &$xy(x^2-y^2)$ & & V \\  
\vspace{.1cm} 
$B_{1g}$  & &$d_{x^2-y^2}$& &  $x^2-y^2$ & & V \\  
\vspace{.1cm}  
$B_{2g}$  & &$d_{xy}$ && $xy$ & & V \\  
\vspace{.1cm}  
$E_{g} \hspace{0.1cm} (1,0)$  & & -  & & $xz$ & & V+H \\ 
\vspace{.1cm}  
$E_{g} \hspace{0.1cm} (1,1)$ & & -  & & $(x+y)z$ & & V+H \\  
\vspace{.4cm}  
$E_{g} \hspace{0.1cm} (1,i)$&  & hybrid & & $(x+iy)z$& & H+points \\ \hline 
\end{tabular*} 
\vglue-3mm 
\end{table} 
%-------------------------------------------- 
Let us now apply Eq.~1 to CeIrIn$_5$. The allowed order parameter representations in tetragonal symmetry \cite{D4h} are listed in Table~I (for singlet pairing). Two line node topologies are possible: {\it vertical} line nodes (where the Fermi surface cuts a vertical plane, e.g. $x=0$), such as in the two $d$-wave states ($d_{x^2-y^2}$ in $B_{1g}$ or $d_{xy}$ in $B_{2g}$), and a {\it horizontal} line node (where the Fermi surface cuts the $z=0$ plane), such as in the hybrid gap of the $E_g~(1,i)$ state. The simplest gap functions are $\Delta = \Delta_0 {\rm cos2\phi}$ and $\Delta = 2 \Delta_0 {\rm cos\theta} {\rm sin\theta}{\rm e}^{i\phi}$, for $d$-wave and $E_{g} (1,i)$ states, respectively.  The corresponding nodal structures are illustrated in Fig.~\ref{fig:gaps}. 
Let us apply Eq.~1 to such a hybrid gap function, for which $a = 3/2$ and $\mu \equiv \mu_{\rm line} = 2$ \cite{Graf-JLTP}. Using the known values of $\gamma_N$ (7300~J~K$^{-2}$~m$^{-3}$ \cite{Movshovich}),  $v_F$ (2$\times$ 10$^{4}$~m/s, in the basal plane \cite{Haga}), and $\Delta_0$ (2.5~$k_B T_c$ \cite{NQR2}), Eq.~1 yields $\kappa_{0a}/T = 28 $~mW/K$^2$~cm. From Fig.~\ref{fig.kT}, the experimental value is $\kappa_{0a} / T \simeq 20$~mW/K$^2$ cm, in remarkable agreement with the theoretical estimate. 
If we model the hybrid gap structure in terms of a single effective slope of the gap at the line node averaged over the various sheets of the Fermi surface of CeIrIn$_5$, that parameter comes out to be $\mu_{\rm eff} = 2.8$.
While this quantitative agreement with theory is compelling confirmation for the presence of a line node in the gap of CeIrIn$_5$, it actually says little about its location. Indeed, the corresponding estimate for a $d$-wave gap gives a similar value for $\kappa_{0a}/T$.
The diagnostic power of thermal conductivity in determining the detailed topology of the gap comes from its {\it directional} character, accessed by sending the current in distinct high-symmetry directions of the crystal. This was not done in the previous heat transport study \cite{Movshovich} and, to the best of our knowledge, no directional measurement of the gap has been reported so far for CeIrIn$_5$.   
 
%------------------- figure 2:  gap structures  
\begin{figure} 
\centerline{ 
\scalebox{0.54}{ 
\includegraphics{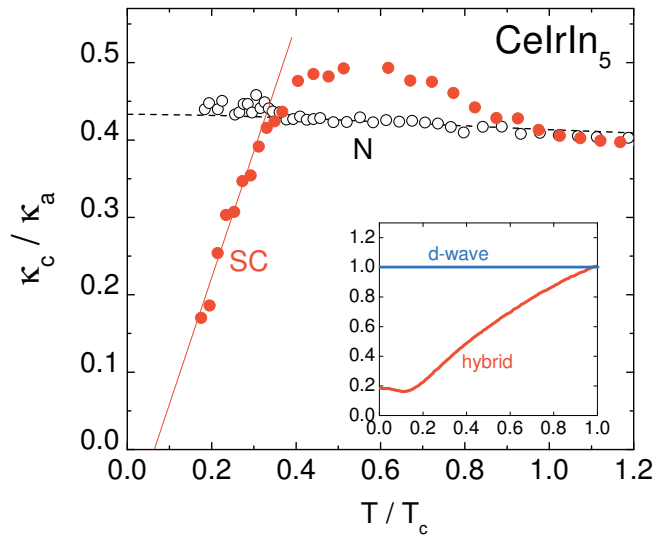}}} 
\caption{\label{fig:kcka} 
Temperature dependence of the anisotropy ratio $\kappa_c/\kappa_a$ of CeIrIn$_5$, in the normal state (N) and in the superconducting state (SC). The dotted line is the ratio of the two fit lines (to the normal state data) displayed in Fig.~1 and the solid line is a linear fit to the superconducting state data below $T_c / 3$. The precipitous drop at low temperature reflects a strongly anisotropic gap whose nodal structure is inconsistent with vertical line nodes (running along the $c$-axis). 
The small peak below $T_c$ is due to inelastic scattering (see text). Inset: calculated anisotropy (normalized at $T_c$) for the gaps shown in Fig.~\ref{fig:gaps}, namely $d$-wave \cite{Vekhterprivit} and hybrid \cite{Fledderjohann}. } 
\vglue-3mm  
\end{figure}

%------------------------------------------------------------------------- 
%%%%%%%%%%%%%%%%  ANISOTROPY 
\textit{Anisotropy}. 
As seen in Fig.~1, applying the current along the $c$~axis reveals a qualitatively different limiting behaviour, whereby $\kappa_c / T \to 0$ as $T \to 0$. Simple $T^2$ or $T^3$ extrapolations yield $\kappa_{0c}/T$ values no greater than 1-2~mW/K$^2$~cm, an order of magnitude smaller than $\kappa_{0a}/T$. 
Fig.~\ref{fig:kcka} shows the anisotropy ratio, $\kappa_c / \kappa_a$, as a function of temperature, in both normal and superconducting states.  
In the normal state, $\kappa_c / \kappa_a$ is virtually independent of temperature, with $\kappa_a / \kappa_c \simeq 2.5$. The anisotropy in $\rho(T)$ is similarly constant, even well beyond the Fermi-liquid $T^2$ regime, with $\rho_c / \rho_a \simeq$~2.7 between 1.2 and 8~K. This simply reflects the anisotropy of the Fermi velocity (or mass tensor). A $T$-independent normal-state anisotropy was also found in UPt$_3$ \cite{Lussier,RMP-UPt3}. 
%---------------------- figure 3: kc/ka Vs. T 
%------------------------------------------------------------------------ 
 
The superconducting state anisotropy is strikingly different, a difference that can only come from gap anisotropy. Two distinct features are manifest: 1) a slight increase immediately below $T_c$ and 2) a precipitous drop below $T \simeq T_c/3$. These two features combine to produce a broad peak centered at $T \simeq T_c / 2$. We attribute the first feature to an {\it anisotropic} suppression of inelastic scattering, brought about as electrons pair up (anisotropically) and cease to participate in the electron-electron scattering responsible for the $bT^2$ term in $\kappa_N / T$.  
 
The second feature is directly diagnostic of the nodal structure, as it comes from low-energy quasiparticles. The factor of $\sim 3$ drop in $\kappa_c / \kappa_a$ between $T_c / 3$ and $T_c / 8$ clearly extrapolates to a very small value as $T \to 0$. This reveals a {\it qualitative} $a$-$c$ anisotropy in the average velocity of thermally excited nodal quasiparticles. In other words, those $k$-states responsible for $c$-axis conduction in the normal state appear to be much more strongly gapped. {\it This excludes any nodal structure for which the line nodes are along the $c$-axis, irrespective of the shape of the Fermi surface}. Indeed, such vertical line nodes would simply reproduce the underlying anisotropy of $v_F$, and $\kappa_c / \kappa_a$ would basically mimic the normal state anisotropy. This expectation, confirmed by calculations \cite{Vekhterprivit}, is illustrated in the inset of Fig.~3 (horizontal blue line). 
 A modulation of the gap maximum along the $c$-axis, whereby $\Delta_0 = \Delta_0(\theta)$, can produce some additional anisotropy in the superconducting state, but this is typically modest and weakly $T$ dependent \cite{Vekhterprivit}. More importantly, it would never bring to nearly zero for $J \parallel c~(\theta=0)$ the residual linear term present for $J \parallel a~(\theta=\pi / 2)$.  
  
By excluding vertical line nodes in the gap of CeIrIn$_5$, our study eliminates all allowed representations for the order parameter, except one: the two-component $E_{g}$ representation (see Table~I). In particular, both $d$-wave states are ruled out: $d_{x^2 - y^2}$ and $d_{xy}$, respectively in $B_{1g}$ and $B_{2g}$ symmetry. Of the three states allowed in the $E_g$ representation, only the $(1,i)$ state is generically free of vertical line nodes. Its typical $(x+iy)z$ dependence produces a hybrid gap, which possesses, in addition to the line node in the basal plane ($z=0$), point nodes along the $z\parallel c$ direction, at $x=y=0$ (see Fig.~2). Note that this state breaks time-reversal symmetry, and will therefore spontaneously generate an internal magnetic moment around impurities. $\mu$SR measurements on CeIrIn$_5$ have not detected such moments \cite{Higemoto}, possibly because the associated fields are too small in these high purity samples, as also found in high quality crystals of UPt$_3$ \cite{Dalmas}, a superconductor for which the accumulated evidence points overwhelmingly to a ground state with broken time-reversal symmetry 
\cite{RMP-UPt3}. 
 
We now consider whether our data is compatible with another special feature of the $E_g~(1,i)$ state: the $c$-axis point nodes of its hybrid gap. These are {\it linear} point nodes, {\it i.e.} $\Delta(\theta) \propto \theta$, such that $N(E) \propto E^2$, which implies that $\kappa_{0} / T$ in the $c$ direction is {\it not} universal. 
Theory shows that $\kappa_{0c} / T$ should  be smaller than $\kappa_{0a} / T$ by a factor $\simeq \sqrt{\hbar \Gamma/\Delta_0} \times \mu_{\rm line} / \mu^2_{\rm point}$ (in the unitary limit) \cite{Graf-JLTP}, where $\Gamma$ is the impurity scattering rate and $\mu_{\rm line}$~($\mu_{\rm point}$) is the slope of the gap at the line (point) node.  
The predicted anisotropy for $\hbar \Gamma = 0.1~k_B T_c$ and $\mu_{\rm line} = \mu_{\rm point} = 2$ is shown in the inset of Fig.~3 (red line; taken from \cite{Fledderjohann}).
While the detailed temperature dependence of the data on this multi-band material is not expected to be captured by the simple model of a single spherical Fermi surface, it is nevertheless meaningful to look at the $T \to 0$ limit, governed entirely by the slope of the gap at the nodes \cite{Graf,Graf-JLTP}.
The calculation shown in Fig.~3 yields a residual anisotropy $\kappa_c/\kappa_a$ that is 20~\% of the normal state anisotropy at $T \to 0$.
This is compatible with the data, where the lowest point only restricts the residual anisotropy ratio to be less than 40~\% of its normal-state value.  
Even a sizable increase in $\Gamma$ would not necessarily invalidate this compatibility since it could easily be compensated by a reduction in the gap parameter $\mu_{\rm point}$.
To summarize, while $\kappa_{0a}/T$ is inconsistent with an $s$-wave gap ($A_{1g}$ in Table~I) and $\kappa_{0c}/T$ is inconsistent with a $d$-wave gap (or any gap with a vertical line node), both $\kappa_{0a}/T$ and $\kappa_{0c}/T$ are quantitatively consistent with a hybrid gap.
%%%%%%%%%%%%% COMPARE TO CeCoIn5   
 
We note that the anisotropy of heat conduction was also measured in CeCoIn$_5$ \cite{Makariy}, but the presence of unpaired electrons in that material produces an unexpectedly large and {\it isotropic} residual linear term which totally masks any anisotropy that might come from the coexisting nodal quasiparticles.  
It should be emphasized that the lack of a sizable residual linear term in the $c$-axis data reported here rules out the possibility of such uncondensed electrons in CeIrIn$_5$. 
 
%%%%%%%%%%%%%%%%%%   CONCLUSIONS 
In conclusion, the in-plane thermal conductivity $\kappa_a$ of CeIrIn$_5$ measured down to $T_c / 8$ confirms unambiguously the presence of line nodes in the superconducting gap. The $c$-axis conductivity $\kappa_c$ reveals a profound anisotropy as $T \to 0$, which rules out the possibility that these line nodes are vertical (along the $c$-axis). This eliminates all but one of the pairing states allowed in $D_{4h}$ symmetry, including the $d$-wave state proposed for the closely related compound CeCoIn$_5$. This leaves as sole candidate for CeIrIn$_5$ the $(1,i)$ state of the $E_g$ representation, also a prime candidate for the superconductor UPt$_3$ \cite{RMP-UPt3}. 
The $T \to 0$ value of $\kappa / T$ in both high-symmetry directions is in good quantitative agreement with theoretical calculations for this state.
This therefore points to superconducting order parameters of different symmetry in the two isostructural members of the Ce$M$In$_5$ family of nearly magnetic heavy-fermion metals.  
It will be interesting to examine how these differences might arise from the respective magnetic fluctuation spectra. Direct experimental confirmation of the presence of $c$-axis point nodes and broken time-reversal symmetry, both implied by the $E_g~(1,i)$ state, is called for. In principle, both should be revealed by doping with impurities.   
%%%%%%%%%%%%%%%%%   ACKNOWLEDGMENTS 
\vspace{0.4cm} 
We are grateful to I.~Vekhter and A. Vorontsov for sharing their calculations before publication, to M.J. Graf for a careful reading of the manuscript, and to B.~Davoudi, C.~Lupien, J.~Paglione, K.~Samokhin, A.-M.~Tremblay, for helpful discussions. This work was supported by the Canadian Institute for Advanced Research and a Canada Research Chair (L.T.), and funded by NSERC of Canada and FQRNT of Quebec. It was partially carried out at the Brookhaven National Laboratory, which is operated for the U.S. Department of Energy by Brookhaven Science Associates (DE-Ac02-98CH10886). 

%%%%%%%%%%%%%%%%%   BIBLIOGRAPHY 


\begin{references} 
 
\bibitem{Mathur} N. D. Mathur {\it et al.}, Nature {\bf 394}, 39 (1998). 
 
\bibitem{Petrovic} C. Petrovic {\it et al.}, J. Phys.: Condens. Matter {\bf 13}, L337 (2001). 
 
\bibitem{Petrovic-Ir} C. Petrovic {\it et al.}, Europhys. Lett. {\bf 53}, 354 (2001). 
 
\bibitem{IzawaCo} K. Izawa {\it et al.}, Phys. Rev. Lett. {\bf 87}, 057002 (2001). 
 
\bibitem{Aoki} H. Aoki {\it et al.}, J. Phys.: Condens. Matter {\bf 16}, L13 (2004). 
 
\bibitem{Vekhter} A. Vorontsov {\it et al.}, Phys. Rev. Lett. {\bf 96}, 237001 (2006). 
 
\bibitem{Takimoto} T. Takimoto {\it et al.}, Phys. Rev. B {\bf 69}, 104504 (2004). 
 
\bibitem{Watanabe} S. Watanabe {\it et al.}, J. Phys. Soc. Jpn {\bf 75}, 043710 (2006). 
 
\bibitem{Tanaka} K. Tanaka {\it et al.}, J. Phys. Soc. Jpn {\bf 75}, 024713 (2006). 
 
\bibitem{Maehira} T. Maehira, {\it et al.}, J. Phys. Soc. Jpn {\bf 72}, 854 (2005). 
 
\bibitem{Haga} Y. Haga {\it et al.}, Phys. Rev. B {\bf 63}, 060503(R) (2001). 
 
\bibitem{Movshovich} R. Movshovich {\it et al.}, Phys. Rev. Lett. {\bf 86}, 5152 (2001).  
 
\bibitem{NQR} Y. Kohori {\it et al.}, Phys. Rev. B {\bf 64}, 134526 (2001). 
 
\bibitem{NQR2} G.Q. Zheng {\it et al.}, Phys. Rev. Lett. {\bf 86}, 4664 (2001). 
 
\bibitem{Nicklas} M. Nicklas {\it et al.}, Phys. Rev. B {\bf 70}, 020505(R) (2004). 
 
\bibitem{Kawasaki} S. Kawasaki {\it et al.}, Phys. Rev. Lett. {\bf 94}, 037007 (2005). 
 
\bibitem{Lussier} B. Lussier {\it et al.}, Phys. Rev. Lett. {\bf 73}, 3294 (1994); B. Lussier {\it et al.}, Phys. Rev. B {\bf 53}, 5145 (1996). 
 
\bibitem{RMP-UPt3} R. Joynt {\it et al.}, Rev. Mod. Phys. {\bf 74}, 235 (2002). 
 
\bibitem{Msmith} M. Smith {\it et al.}, Phys. Rev. B {\bf 71}, 014506 (2005).  
 
\bibitem{Makariy} M. A. Tanatar {\it et al.}, Phys. Rev. Lett. {\bf 95}, 067002 (2005). 
 
\bibitem{Fledderjohann} A. Fledderjohann {\it et al.},  Solid State Comm. {\bf 94}, 163 (1995). 
 
\bibitem{Graf} M. J. Graf {\it et al.}, Phys. Rev. B {\bf 53}, 15147 (1996). 
 
\bibitem{Norman} M.R. Norman {\it et al.},  Phys. Rev. B {\bf 53}, 5706 (1996). 
\bibitem{Graf-JLTP} M. J. Graf {\it et al.}, J. Low Temp. Phys. {\bf 102}, 367 (1996). 
 
\bibitem{Taillefer} L. Taillefer {\it et al.}, Phys. Rev. Lett. {\bf 79}, 483 (1997); M. Chiao {\it et al.},  Phys. Rev. B {\bf 62}, 3554 (2000). 
 
\bibitem{Suzuki} M. Suzuki {\it et al.}, Phys. Rev. Lett. {\bf 88}, 227004 (2002). 
 
\bibitem{D4h} C.C. Tsuei {\it et al.}, Rev. Mod. Phys. {\bf 72}, 969 (2000). 
 
\bibitem{Vekhterprivit} I.~Vekhter and A. Vorontsov, unpublished. 
 
\bibitem{Higemoto} W. Higemoto {\it et al.}, J. Phys. Soc. Jpn {\bf 71}, 1023 (2002). 
\bibitem{Dalmas} P. Dalmas de Reotier {\it et al.}, Phys. Lett. A {\bf 205}, 239 (1995). 
 
\end{references}
\end{document}